# Detectability of Surface Biosignatures for Directly-Imaged Rocky Exoplanets

Schuyler R. Borges[1,2], Gabrielle G. Jones[1,2], Tyler D. Robinson[1,2,3,4]

## Abstract

Modeling the detection of life has never been more opportune. With next generation space telescopes, like the currently developing Habitable Worlds Observatory (HWO) concept, we will begin to characterize rocky exoplanets potentially similar to Earth. However, currently, few realistic planetary spectra containing surface biosignatures have been paired with direct imaging telescope instrument models. Therefore, we use a HWO instrument noise model to assess the detection of surface biosignatures affiliated with oxygenic, anoxygenic, and nonphotosynthetic extremophiles. We pair the HWO telescope model to a 1-D radiative transfer model to estimate the required exposure times necessary for detecting each biosignature on planets with global microbial coverage and varying atmospheric water vapor concentrations. For modeled planets with 0% - 50% cloud coverage, we determine pigments and the red edge could be detected within 1,000 hours (100 hours) at distances within 15 pc (11 pc). However, tighter telescope inner working angles (2.5 $\lambda/D$) would allow surface biosignature detection at further distances. Anoxygenic photosynthetic biosignatures could also be more easily detectable than nonphotosynthetic pigments and the photosynthetic red edge when compared against a false positive iron oxide slope. Future life detection missions should evaluate the influence of false positives on the detection of multiple surface biosignatures. **Keywords:** Extremophiles--Pigments--Red edge--False positive--Telescope--Reflectance Spectroscopy.

[1] Department of Astronomy and Planetary Science, Northern Arizona University, USA
[2] Habitability, Atmospheres, and Biosignatures Laboratory, University of Arizona, USA
[3] Lunar and Planetary Laboratory, University of Arizona, USA
[4] NASA Nexus for Exoplanet System Science Virtual Planetary Laboratory, USA

Corresponding author: Schuyler R. Borges, email: srb558@nau.edu

1. **Introduction**

A key step toward the discovery of life on a distant planet is to understand the types of biosignatures that can be remotely identified. An increase in reflectivity of Earth's planetary spectrum between red (0.7 µm) and near-infrared (NIR; 0.8 - 2.5 µm) wavelengths (Sagan *et al.*, 1993; Cowan *et al.*, 2009) is called the "vegetation red edge" (Gates *et al.*, 1965; Knipling, 1970; Seager *et al.*, 2005). This red edge is attributed to oxygenic photosynthetic life absorbing and scattering energy at the wavelength ranges where maximum energy is emitted by our stellar host (Kiang *et al.*, 2007a; Kiang *et al.*, 2007b). While oxygenic photosynthetic life has dominated Earth's atmospheric spectrum for hundreds of millions of years (O'Malley-James and Kaltenegger, 2018), other forms of life can also contribute to a planetary spectrum. Nonphotosynthetic microbes contain pigments, which can absorb energy at ultra-violet (UV) and visible wavelength ranges functionally independent of the stellar host (Schwieterman *et al.*, 2015). Biological influences on planetary reflectance spectra are a key motivation for the development of exoplanet direct imaging mission concepts, including the recently-studied Habitable Exoplanet Observatory (HabEx; Gaudi *et al.*, 2018) and the Large UltraViolet-Optical-InfraRed Surveyor (LUVOIR; Roberge and Moustakas, 2018). With the recent prioritization of exo-Earth direct imaging in the 2020 Decadal Survey on Astronomy and Astrophysics (National Academies of Sciences, Engineering, and Medicine, 2021) — with a mission concept now dubbed the Habitable Worlds Observatory (HWO) — understanding how surface biosignatures could be detected on exoplanets has never been more critical.

Multiple studies have used modeling to predict which spectral signatures could serve to identify potentially habitable planets (Hegde and Kaltenegger, 2013; Sanromá *et al.*, 2013; Sanromá *et al.*, 2014; O'Malley-James and Kaltenegger, 2018). The influence of surface vegetation cover (*i.e.* mosses, ferns, trees, and cacti) on Earth's reflectance spectrum has been modeled over Earth's last 500 million years to understand the variation of the vegetation red edge as a spectral signature over time (Arnold *et al.*, 2009; O'Malley-James and Kaltenegger, 2018). To discriminate between potential Earth-like exoplanets with dominant deserts, land vegetation, and microbial mats, four different continental environments were also modeled for Earth 500 million years ago (Sanromá *et al.*, 2013). Additionally, Earth around 3 billion years ago has been compared to a modern Earth to determine that a relatively young Earth-like planet, covered with purple anoxygenic photosynthetic bacteria, could be distinguished from an older Earth with more modern life as we know it (Sanromá *et al.*, 2014). A range of reflectance spectra of potentially habitable exoplanets with various environmental conditions have also been modeled to determine the detectability of extremophilic microorganisms on their surface (Hegde and Kaltenegger, 2013). Extreme environmental planetary conditions can be influenced by proximity to stellar hosts, presence of an atmosphere, and presence of liquid solvents. A unique spectral archive of extremophilic organisms has been developed specifically for these types of studies (Hegde *et al.*, 2015). Within some of these spectra are nonphotosynthetic pigments, which may be detectable surface biosignatures on rocky exoplanets (Schwieterman *et al.*, 2015). Overall, spectra of microorganisms and vegetation have been used in modeling efforts for a range of planetary environments to determine the detectability of biosignatures on habitable exoplanets.

Most studies modeling the detection of various habitable world spectra have used filter photometry (Sanromá *et al.*, 2013; Hegde and Kaltenegger, 2013; Sanromá *et al.*, 2014; Schwieterman *et al.*, 2015; O'Malley-James and Kaltenegger, 2018); however, future space-based telescope instrument observation simulation models are critical for better estimating biosignature detectability (Lustig-Yaeger *et al.*, 2018; Feng *et al.*, 2018; Wang *et al.*, 2018). For instance, atmospheric biosignature gases of Earth could be detected using HabEx and LUVOIR models (Feng *et al.*, 2018; Wang *et al.*, 2018), and the red edge, present in an Earth-like exoplanet spectrum, could be retrieved with any future direct-imaging telescope observing reflected light (Gomez Barrientos *et al.*, 2023). Conversely, further studies of these biosignatures can more importantly inform the design of future space telescopes. Wang *et al.* (2018) determined the required spectral resolution, starlight suppression, and exposure time necessary to detect atmospheric biosignatures on an Earth-like exoplanet using HabEx and LUVOIR telescope models. In doing so, these authors ultimately provide baseline requirements for these telescopes' designs. Determining the detection limits and understanding the characteristics of a variety of biosignatures, including surface biosignatures, will further inform the design of a future HWO mission and increase chances of detecting life on another world.

Accounting for potential false positive detections of life can also be an important part of these modeling efforts. Reflectance spectra of cinnabar and sulfur have slopes reminiscent of the photosynthetic red edge (Seager *et al.*, 2005; Schwieterman *et al.*, 2018). An increase in spectral slope in the NIR of iron oxides can also act as a false positive for the red edge (Sparks *et al.*, 2009; Kokaly *et al.*, 2017). While these minerals are known to have spectral features similar to the red edge, none have been directly incorporated into habitable world spectra and their detection models. Accounting for false positives in these models will rule out uncertainty in, and support the confirmation of, a detection of a red edge slope. Therefore, modeling efforts should not only account for abiotic spectra, but more specifically false positive spectra when considering future remote observations from space-based telescopes.

In this study, we investigate the feasibility of using an HWO-like space-based telescope to detect both the red edge and pigment surface biosignatures for rocky exoplanets, while considering the influence of false positives. Our models adopt oxygenic photosynthetic, anoxygenic photosynthetic, and nonphotosynthetic organismal spectra from Hegde *et al.* (2015) and Borges *et al.* (2023) spectral archives and consider potential false positive effects from spectral slopes analogous to those created by iron oxide minerals. Planetary spectra are predicted using a 1-D radiative transfer model, and observational noise effects are incorporated using a direct imaging instrument model. Using our model, we investigate the difference in integration time necessary for detecting (1) anoxygenic and nonphotosynthetic pigments against an abiotic background and (2) the red edge against a false positive iron oxide feature obtained from Kokaly *et al.* (2017). Our results comprehensively show the utility in pairing exoplanet space-based direct-imaging noise models to assess the detectability of surface life on a distant Earth-like world.

**2. Methods**

Here, we describe the microbial communities we chose to represent extremophilic life that could exist on different types of desert-like worlds. We obtain their reflectance spectra collected in the field and lab to identify regions containing biosignatures, such as pigments and the photosynthetic red edge. We generate spectra of the microbial communities with and without these biological signatures in order to find the time it would take HWO to differentiate between the two. We consider the influence of false positives on the detection of the red edge by replacing the red edge with a mineralogic false positive in the reflectance spectra of oxygenic photosynthetic microbial communities. We detail the inclusion of these microorganisms' reflectance in the production of top-of-atmosphere planetary spectra using atmospheric radiative transfer and noise models, which simulate HWO observations. We determine the required exposure times necessary for detecting the surface biosignatures affiliated with our microbial life.

*2.1 Choosing Microbial Spectra*

While current and recent past environments on Earth have hosted complex life (*e.g.* plants, trees, forests), it may be more likely to find single-cell life, like microorganisms, on another world ([Catling *et al.*, 2005](#)). A number of rocky exoplanets could exist at the limits of habitable zones and be extremely water-limited ([Abe *et al.*, 2011](#); [O'Malley-James and Kaltenegger, 2018](#)). Thus, extremophilic microbial life could exist on a number of these more desert-like rocky exoplanets ([Hegde and Kaltenegger, 2013](#); [Hegde *et al.*, 2015](#); [O'Malley-James and Kaltenegger, 2018](#)).

To represent a diversity of potential life on these desert-like rocky exoplanets, we chose extremophilic microbial communities from three types of desert environments: 1) cold and dry, 2) hot and dry, and 3) hot and wet (acidic saline pools in a hot and dry desert). Cold and dry deserts, as well as hot and dry deserts, have extreme temperatures, high aridities, and high ultra-violet (UV) radiation. Despite usually being water-limited, deserts can have small ponds or lakes with high salinities as well ([Pinti, 2011](#)). The temperature, salinity, pH, and overall chemistry in these environments differ, so microbial communities within these deserts also vary. We chose oxygenic photosynthetic, anoxygenic photosynthetic, and nonphotosynthetic microbial communities from these three different types of desert environments to represent differences in feasibility and efficiency of detecting a range of surface biosignatures. Largest spectral differences amongst microbes and their various red edge and pigmented compositions were also chosen to effectively represent a range of potential surface biosignatures.

Representing life on a cold and dry planet, we chose oxygenic photosynthetic microbial communities, black and orange microbial mats, from Crescent and Canada Streams in the Fryxell Basin of Taylor Valley, Antarctica (**Fig. 1**). Black microbial mat predominantly consists of *Nostoc*, and its spectrum has a slight chlorophyll-*a* absorption feature at 0.675 µm and a broad, lower sloping photosynthetic red edge from ~0.7 - 1.0 µm, which is influenced by the abundant photoprotective pigment, scytonemin (**Fig. 2**; Garcia-Pichel and Castenholz, 1991; Vincent *et al.*, 1993b; Alger *et al.*,1997). Orange microbial mat is predominantly composed of *Phormidium* and *Oscillatoria*, and its spectrum is composed of carotenoid absorptions from 0.4 - 0.55 µm, a phycocyanin absorption at 0.62 µm, a chlorophyll-*a* absorption at 0.675 µm, a very steep red edge from ~0.7 - 0.75 µm, and a potential water absorption around 0.975 µm (**Fig. 2**; Vincent *et al.*, 1993a; Alger *et al.*, 1997). While these microbial communities have numerous biosignatures, for the purpose of our study, we only highlight their red edges. We define the low reflectance side of their red edge as the "foot," and the high reflectance side of their red edge as the "shoulder" (**Fig. 2**). Both types of microbial mats are well-adapted to the radiation, cold temperature, and aridity of the cold and Mars-like deserts of the McMurdo Dry Valleys.

We chose a nonphotosynthetic organism from the genus *Arthrobacter* to represent a hot and dry desert-like world (**Fig. 1**). This heterotroph was sampled from the Atacama Desert in Chile, and *Arthrobacter* sp. is known to survive the arid, high UV radiation environment of the Atacama, which is an analog environment to Mars (Azua-Bustos *et al.*, 2012; Dsouza *et al.*, 2015). The pigments that contribute to its unique spectral features span a range of hues: orange and yellow pigments include riboflavin and carotenoids, blue and green pigments include indigoidine, indochrome, and derived salts, and red pigments include porphyrins and carotenoids (Sutthiwong et al., 2014). *Arthrobacter* sp. from Hegde *et al.*, 2015 appears to have an orange-yellow color, which could indicate the presence of riboflavin and carotenoids (Hegde *et al.*, 2015). However, further analysis on this organism cultured from the Atacama Desert would be needed to definitively determine its pigment composition. Regardless, riboflavin and carotenoids perform a range of functions in the UV and have absorption features up to ~0.58 μm (**Fig. 2**; Sutthiwong et al., 2014). While riboflavin absorption maxima change depending on pH and solvent, generally riboflavin has absorption peaks anywhere from 0.44 – 0.447 μm and at 532 μm in the visible wavelength range (Drössler *et al.*, 2002; Sikorska *et al.*, 2005; Zirak *et al.*, 2009; Orlowska *et al.*, 2013). However, the *Arthrobacter* sp. spectrum presented here has absorption features at roughly 0.4 μm, 0.47 μm, 0.5 μm, and 0.54 μm, which indicates they are more likely carotenoid absorption features (**Fig. 2**).

For our hot and wet desert-like life, we chose an anoxygenic photosynthetic microbe, *Ectothiorhodospira* sp. str. BSL-9, which is a purple sulfur bacterium found in the anoxic layer of the hypersaline and alkaline Big Soda Lake, Nevada (**Fig. 1**). The reflectance spectrum of *Ectothiorhodospira* sp. str. BSL-9 is dominated by bacteriochlorophyll-*a* at 0.80 μm and 0.85 μm and carotenoids at 0.59 μm, 0.5 μm, and 0.4 μm (**Fig 2**; McCann *et al.*, 2016). This strain of *Ectothiorhodospira* uses arsenic as an electron donor in photosynthesis (McCann *et al.*, 2016), and Archean Earth (4.5 - 2.5 billion years) could have had organisms with similar metabolisms (Sforna *et al.*, 2014). Additionally, hypersaline, anoxic, and perchlorate environments have existed on Mars, and soda lakes have been considered analogs for potential Mars-like environments (Kempe and Kazmierczak, 1997; Mormile *et al.*, 2009; Matsubara *et al.*, 2017). Thus, this microbe could represent potential life on an Archean or early Mars-like exoplanet.

While biomolecules are surface biosignatures that could also be present in these spectra, they absorb wavelengths of light in the near-infrared from 0.8 – 2.4 μm where water and atmospheric gases are also absorbing (Poch *et al.*, 2017; Schwieterman *et al.*, 2018). Therefore, biomolecule spectral features could get obscured by water and atmospheric gas features. Additionally, the spectral range for our HWO-like telescope model ends around 1.03 μm, which means any absorption features beyond this wavelength will not be detectable (see section 2.4). Therefore, our study focuses on the red edge, which occurs around 0.7 μm and pigments, which are present from 0.4 – 0.8 μm (Poch *et al.*, 2017; Schwieterman *et al.*, 2018).

*2.2 Removing Biosignatures from Microbial Spectra*

The detectability of a spectral feature is defined by the integration time required to detect a feature of interest by its difference from a feature-less smooth continuum (Robinson *et al.*, 2016). In other habitable world models, an abiotic reference, often soil or geologic mineral, is compared with biologic world spectra instead (Hegde and Kaltenegger, 2013; Sanromá *et al.*, 2013; Schwieterman *et al.*, 2015). Since we are interested in the specific detection of pigments and the photosynthetic red edge in our reflectance spectra, we isolate the spectral region of each biosignature and replace the biosignature with a linear interpolation — or structureless, smooth continuum — between the lips of the reflectance well of a biosignature feature. In the case of the red edge, which is known to have false positives (Seager *et al.*, 2005; Sparks *et al.*, 2009; Kokaly *et al.*, 2017; Schwieterman *et al.*, 2018), this linear interpolation is the spectral slope of a geologic mineral, which could be mistaken for the red edge, irrespective of wavelength. By comparing the spectral differences between the biotic and abiotic surfaces, the distinguishability can be quantified.

To imitate an abiotic spectrum of black and orange microbial mat, we replaced the photosynthetic red edge with a false positive red edge slope from iron oxide and hydroxide (**Fig. 3**; **Fig. 4**). While cinnabar and sulfur contain spectral slopes similar to that of the red edge (Seager *et al.*, 2005; Schwieterman *et al.*, 2018), we use spectral slopes from iron oxide and hydroxide because iron is likely to be more common on rocky exoplanets, given their expected iron composition (Howard *et al.*, 2013). Iron oxide and hydroxide spectra were collected from the Kokaly *et al.* (2017) spectral database (**Fig. 3**). Linear fits were calculated for each iron mineral slope and then averaged (**Fig. 3**). This averaged slope was then substituted in place of the photosynthetic red edge, which was at different wavelengths in black and orange microbial mat spectra (**Fig. 4**). Differences between the abiotic and biotic spectrum for black microbial mat were not as large as those between the abiotic and biotic spectrum for orange microbial mat (**Fig. 4**).

Unlike the red edge, anoxygenic and nonphotosynthetic pigments do not have any known false positives (Harman and Domagal-Goldman, 2018). So, we created abiotic spectra for *Arthrobacter* sp. and *Ectothiorhodospira* sp. str. BSL-9 by linearly interpolating between local maxima of pigment absorption features (**Fig. 4**). For *Arthrobacter* sp., points were linearly interpolated between 0.4 μm and 0.58 μm (**Fig. 4**). For *Ectothiorhodospira* sp. str. BSL-9, maximum reflectance points were linearly interpolated at around 0.4 μm, 0.65 μm, and 0.95 μm (**Fig. 4**).

*2.3 Atmospheric Modeling of Simulated Planets*

As is the case on Earth, Earth-like planets are likely to have an atmosphere, so contributions from atmospheric absorption and scattering will influence the planetary spectrum alongside surface reflectance. To account for an atmosphere, we used the 1-D (vertical) Spectral Mapping Atmospheric Radiative Transfer (SMART) model (developed by D. Crisp; [Meadows and Crisp, 1996](#)), which generates top-of-atmosphere planetary radiance spectra. We assumed an Earth-like atmospheric structure (temperature, pressure, and gas mixing ratios) and atmospheric opacities calculated using HITRAN 2016 ([Gordon *et al.*, 2016](#)). Biotic and abiotic reflectance spectra created in 2.2 were input as surface reflectances in the SMART model, assuming fully global surface coverage of our microbial communities for optimal detection.

Since the surface life we are simulating is associated with desert environments, we created two planetary atmospheric scenarios: 1) the planet has the same amount of atmospheric water vapor as Earth and 2) the planet has 10% of Earth's atmospheric water vapor concentration. In the first case, we assume the presence and absence of varying cloud coverage. Clouds are known to scatter light, affecting the overall planetary spectrum and making it more challenging to detect biosignatures ([Tinetti *et al.*, 2006](#); [Sanromá, *et al.*, 2013](#); [Sanromá, *et al.*, 2014](#); [Schwieterman, *et al.*, 2015](#); [O'Malley-James and Kaltenegger, 2018](#)). A partially clouded planetary spectrum can be calculated as the linear combination of a planetary spectrum without cloud cover, with optically-thick low clouds, and with optically-thick high clouds ([Kaltenegger *et al.,* 2007](#); [King *et al.*, 2013](#)). For both biotic and abiotic surfaces, we generated SMART top-of-atmosphere radiance at these three cloud conditions and calculated reflectance. We then linearly mixed these three SMART-generated spectra to produce planetary spectra with 0%, 25%, 50%, and 75% cloud cover for each biotic and abiotic surface. In our second atmospheric case, we only generated top-of-atmosphere reflectance for both biotic and abiotic surfaces with no cloud cover, assuming 10% of Earth's atmospheric water vapor would prevent the production of clouds. These twenty top-of-atmosphere reflectance spectra were then input into the HWO instrument model and compared with the same twenty spectra with abiotic surface compositions (**Table 1**).

*2.4 HWO Instrument Noise Model*

A number of instrumental and astrophysical noise sources can interfere with the detection of planetary spectra obtained from telescopes ([Stark *et al.*, 2014](#)). These include everything from instrument dark current to exozodiacal dust within the planetary system. Therefore, we used a HWO instrument model to account for noise. The overall instrument model used a planetary spectrum and wavelength-dependent noise terms to determine the wavelength-dependent signal-to-noise (S/N) ratios for a standard exposure time. Included noise sources were instrument-related terms, zodiacal light, exozodiacal light, and stellar leakage through the high contrast imaging system. Our model was adapted from the "coronagraph" open source python package created by [Lustig-Yaeger *et al.* (2019)](#) and incorporated an original noise model script developed by T. Robinson ([Robinson *et al.*, 2016](#)). Noise counts were calculated using methods outlined in [Robinson *et al.* (2016)](#). Instrument parameters were based on most recent estimates of the HabEx and LUVOIR space-telescopes (**Table 2**). We assumed the planet had an Earth-radius and semi-major axis (1 AU) and that the planet orbited a Sun-like star (1 solar-radius and temperature of 5,780 K). Our adopted phase angle was 90° (*i.e.*, quadrature), and distance-dependent results were scaled from a baseline simulation executed at a target distance of 5 pc. We varied the inner working angle (IWA) from 3.5 to 2.5 $\lambda/D$ for 0% cloud cover for Earth-like and desert-like atmospheric water vapor planets and 50% cloud cover for Earth-like atmospheric water vapor planets.

Following previous studies that derive a single requisite exposure time for detection from spectrally-varying signal and noise terms (Louie *et al.*, 2018; Lustig-Yaeger et al., 2019; Gialluca *et al.*, 2022), we write the spectrally-dependent signal-to-noise ratio ($S_\lambda/N_\lambda$) as,

$$\frac{S_\lambda}{N_\lambda} = \frac{c_p \delta}{\sqrt{c_p + 2c_b}} \Delta t_{exp}^{1/2}, \quad (1)$$

where $c_p$ is the wavelength-dependent planetary count rate on the detector, $c_b$ is the wavelength-dependent total background count rate on the detector, and $\Delta t_{exp}$ is an exposure time. Here, δ quantifies the strength of the signal we aim to detect, which is taken as the difference in planetary reflectivity between a model with a biotic surface reflectance spectrum versus a model with an abiotic surface reflectance spectrum (*e.g.,* **Fig. 5**). Assume the spectral noise is uncorrelated, the wavelength-integrated signal-to-noise ratio is then,

$$\frac{S}{N} = \sqrt{\sum_\lambda \left(\frac{S_\lambda}{N_\lambda}\right)^2}. \quad (2)$$

Fixing the detection confidence (*e.g.*, setting the spectrally-integrated signal-to-noise ratio equal to 5, which we use throughout and represents a confident "5 sigma" detection) then enables the determination of the exposure time required to achieve this detection.

We input each biotic and abiotic SMART-generated top-of-atmosphere reflectance spectrum into our HWO model and determined the required exposure time necessary for detecting the biosignature. In doing so, we can compare the length of time needed to detect the red edge and pigments.

## 3. Results

*3.1 Top-of-atmosphere reflectance spectra*

For our planetary body with Earth-like water vapor concentrations, the surface reflectance of each microbial community affected the overall shape of the top-of-atmosphere reflectance spectra as a result of surface biosignatures (**Fig. 5**). Atmospheric gas absorption features are known to influence top-of-atmosphere reflectance spectra (Seager *et al.*, 2005; **Fig. 5**). Despite the presence of these gases, the photosynthetic red edge and pigments are still observable in top-of-atmosphere planetary reflectance. Around the presence of their red edges, black and orange microbial mat top-of-atmosphere reflectance spectra have most pronounced features at longer wavelengths between 0.83 - 1.03 µm and 0.64 - 0.81 µm (**Fig. 5**). Pigments, present at shorter wavelengths, also altered the top-of-atmosphere reflectance spectrum to create distinct spectral features (**Fig. 5**). Differences between biotic and abiotic top-of-atmosphere spectra for *Arthrobacter* sp. were pronounced between 0.4 - 0.58 µm (**Fig. 5**). However, *Ectothiorhodospira* sp. str. BSL-9 had a unique spectrum as it had large differences at both short and long wavelengths. Despite the presence of atmospheric gas absorptions, pigments, alongside the red edge, can be distinguished in planetary top-of-atmosphere reflectance spectra.

Our telescope noise model detected the presence of biosignatures in top-of-atmosphere reflectance spectra. As a proof of concept, **Figure 6** shows synthetic telescope observations of the biotic and abiotic spectrum for *Ectothiorhodospira* sp. str. BSL-9 at clear sky conditions, assuming 100 hours of exposure time and a planetary distance of 7.5 pc. The biotic and abiotic case are distinguished insofar as the spectral error bars do not overlap in key regions of the observations indicative of the presence of pigments. As such, these results support our methods of determining the required exposure times by comparing our two different microbial spectra (*i.e.,* biotic and abiotic) at wavelengths where biosignatures are present.

An increase in cloud cover had significant effects on the influence of surface biosignatures on the top-of-atmosphere reflectance spectra. For all microbial communities, the planet's albedo increased with cloud cover (**Fig. 7**). Black microbial mat and *Ectothiorhodospira* sp. str. BSL-9 had the lowest reflectance at 0% cloud cover (**Fig. 7**), but both almost tripled at 75% cloud cover. The difference between all biotic and abiotic spectra also decreased with an increase in cloud cover (**Fig. 7**). Ultimately, cloud cover not only increased the reflectance of planetary spectra, but it also changed the overall shape of the spectra, reducing the magnitude of surface biosignature influence on planetary spectra.

When atmospheric water vapor decreased at 0% cloud coverage, spectral absorption features diminished in both biotic and abiotic planetary spectra (**Fig. 8**). From 0.58 µm onward, a difference in absorption depth between desert-like and Earth-like spectra can be noted in all microbial spectra, especially affecting the region of the red edge in black and orange microbial mat (**Fig. 8**). In contrast, this effect is less pronounced across the wavelength region of pigments in *Ectothiorhodospira* sp. str. BSL-9 and *Arthrobacter* sp. (**Fig. 8**). In fact, biotic and water effects occur at different wavelength regions in *Arthrobacter* sp. spectra with biosignatures influencing shorter wavelengths up to 0.58 µm and water absorption features influencing wavelengths beyond 0.73 µm (**Fig. 5; Fig. 8**). Therefore, atmospheric water vapor does not have an influence on the wavelength region of pigments in *Arthrobacter* sp. (**Fig. 8**). Due to water absorption features being present at red and near-infrared (NIR) wavelengths, differences in atmospheric water vapor will most impact surface biosignatures present at these longer wavelengths.

*3.2 Required exposure times for detecting biosignatures*

Required exposure times were calculated using Equations 1 and 2 and represent the time needed to distinguish our biotic and abiotic spectra. On a planet with an Earth-like atmospheric water vapor concentration, all organisms would require between 7.5 - 460 hours of exposure time to detect (**Table 3**). With a desert-like atmospheric water vapor concentration, between 11 - 93 hours of exposure time would be required to detect these organisms (**Table 4**). All of these estimated exposure times are within the amount of hours typically used for imaging parts of the sky (<1,000 hours; Williams *et al.*, 1996; Beckwith *et al.*, 2006).

Our anoxygenic photosynthetic microbe was fastest to detect. At all cloud covers with an Earth-like concentration of water vapor, black microbial mat required hundreds of hours to detect (**Table 3**; **Fig. 9**), while the second slowest microbe to detect, *Arthrobacter* sp., required ≤ 100 hours to detect (**Table 3**; **Fig. 9**). Orange microbial mat was second fastest to detect, and *Ectothiorhodospira* sp. str. BSL-9 was fastest with its exposure time under 45 hours (**Table 3**; **Fig. 9**). At 0% cloud cover, orange microbial mat required only ~20% more exposure time than *Ectothiorhodospira* sp. str. BSL-9. However, at ≥ 25% cloud cover, orange microbial mat required five times or more exposure time than *Ectothiorhodospira* sp. str. BSL-9, indicating that an increase in cloud coverage also increased how much more detectable *Ectothiorhodospira* sp. str. BSL-9 was then all other microbial communities.

An increase in cloud coverage most often resulted in an increase in exposure time needed to detect black and orange microbial mat, but for *Arthrobacter* sp. and *Ectothiorhodospira* sp. str. BSL-9, low cloud conditions decreased the exposure time. Black microbial mat's fastest detection time was at 0% cloud cover, and while orange microbial mat's fastest time was at 25%, this exposure time was about 2% faster than that at 0% cloud cover, which is a minimal difference. For instance, exposure times for *Arthrobacter* sp. decreased by 32% from 0% to 25% cloud cover. In contrast to black microbial mat, *Ectothiorhodospira* sp. str. BSL-9 took the longest to detect at 0% cloud cover, and the exposure times needed to detect this microbe were fastest at 50% cloud coverage. Exposure times for *Ectothiorhodospira* sp. str. BSL-9 decreased by a factor of four between 0% and 25% cloud cover and three fourths between 25% and 50% cloud cover.

*Ectothiorhodospira* sp. str. BSL-9 continued to be the fastest microbe to detect in an atmosphere with 10% of Earth's water vapor (**Table 4**). However, the overall exposure time decreased by about three fourths in comparison to the exposure time with an Earth-like concentration of Earth's water vapor (**Table 3**; **Table 4**). Orange microbial mat also continued to be second fastest and was 26% faster with a desert-like atmospheric water vapor concentration (**Table 3**; **Table 4**). However, black microbial mat required ~27% less exposure time than *Arthrobacter* sp., as its exposure time decreased by a factor of two and the exposure time for *Arthrobacter* sp. decreased by 1% with the change in atmospheric water vapor (**Table 3**; **Table 4**). This minimal change in exposure time for *Arthrobacter* sp. suggests that spectra of nonphotosynthetic organisms, whose biosignatures absorb at shorter wavelengths than water, are not significantly influenced by the presence of water, while spectra of photosynthetic organisms are.

*3.3 Influence on HWO telescope parameters*

The ability to detect our three microbial communities was influenced by our telescope design. Thus, we determined the relationship between the 1) exposure time, 2) distance to a planetary system, 3) telescope's inner working angle (IWA), and 4) longest wavelength accessible for observation (longest observable wavelength).

With an IWA of 3.5 λ/*D*, required exposure times for detecting all organisms reached a detection limit at a given planetary distance and wavelength (**Fig. 10**). For planets with an Earth-like atmospheric water vapor concentration, the required exposure time to detect the red edge affiliated with black microbial mat became unreasonable (>1,000 hours) at ~8 pc and 1.0 µm, whereas at desert-like atmospheric water vapor concentrations, the maximum planetary distance and observable wavelength were ~9 pc and 0.95 µm (**Fig. 10**). In contrast, orange microbial mat at both atmospheric scenarios became unreasonable (>1,000 hours) to detect at ~11 pc and 0.75 µm. Similarly, at 0% cloud conditions for both atmospheric water vapor conditions, *Arthrobacter* sp. and *Ectothiorhodospira* sp. str. BSL-9 became unreasonable to detect at 12 pc and 0.7 µm. However, with 50% cloud cover at Earth-like atmospheric water vapor conditions, *Arthrobacter* sp. was detectable up to 13 pc and 0.65 µm, whereas *Ectothiorhodospira* sp. str. BSL-9 was detectable up to 15 pc and <0.6 µm (**Fig. 10**). The shoulder of the red edge in black and orange microbial mat as well as the NIR shoulder of the longest wavelength pigment absorptions in the spectrum of *Ectothiorhodospira* sp. str. BSL-9 (grey regions in **Fig. 4**) extend to longer wavelengths than the longest observable wavelengths reported here. However, other defining features of the red edges and pigments are still present at shorter wavelengths than these calculated longest observable wavelengths (**Fig. 4**). Therefore, every surface biosignature would still be detectable at these longest observable wavelengths.

The maximum planetary distance for detecting all organisms increased with an IWA of 2.5 λ/*D*, and the wavelength regions where biosignatures are present remained observable at these farther distances as well. The required exposure time to detect black microbial mat with an Earth-like atmospheric water vapor concentration became unreasonable (>1,000 hours) at ~10 pc and ~1.2 µm, whereas the required exposure time for detecting black microbial mat with desert-like atmospheric water vapor became unreasonable around 11 pc and 1.0 µm (**Fig. 10**). The maximum planetary distance and wavelength for detecting orange microbial mat with 0% and 50% cloud cover at both atmospheric water vapor conditions was ~14 pc and 0.85 µm, and this microbial community's required exposure time was very similar to *Ectothiorhodospira* sp. str. BSL-9 at 0% cloud cover with Earth-like and desert-like atmospheric water vapor conditions (**Fig. 10**). Required exposure times and the maximum planetary distance for detecting *Arthrobacter* sp. remained similar with both IWA, but the longest observable wavelength was 0.95 µm with an IWA of 2.5 $\lambda/D$ and 0.7 µm with an IWA of 3.5 $\lambda/D$. Overall, as the IWA decreased, the exposure time needed for the detection of most surface biosignatures became shorter for planets further away, and longer wavelengths where many studied organisms have red edge features remained accessible as well (**Fig. 10**).

## 4. Discussion

### 4.1 Model Assumptions

Knowledge of the optical properties of various microbial communities will be important for better estimating the exposure times for detecting various surface biosignatures. When creating abiotic surface spectra, biosignature spectral features were smoothed out by linear interpolations in wavelength regions where pigments are thought to absorb and cell structure scatter light (in the case of the photosynthetic red edge). The same grey regions between **Figure 4** and **Figure 5** indicate that the wavelength range of the linear interpolation is ultimately defining the change in top-of-atmosphere reflectance spectra and, thus, the exposure times represented in **Tables 2** and **3**. False positive analogs have not yet been fully explored for biological pigments, and carotenoid pigments will be extremely important for positively detecting life. Alongside this, parameterized treatments for pigments could be incorporated into surface retrievals within inverse models for exoplanets (Gomez Barrientos *et al.*, 2023). More work understanding the biophysical properties and spectral characteristics of these particular surface biosignatures will be important for future life detection missions. For instance, these quantifications of spectral and biological relationships will be necessary for better subjectively and objectively defining abiotic spectra for direct comparisons in telescope models.

While we assumed global coverage of microorganisms on these simulated planets, the possible surface extent of biology on an exoplanet is unknown. Most studies include variations in surface coverage of both biotic and abiotic materials to model realistic planets or worlds similar to Earth (Sanromá *et al.*, 2013; Hegde and Kaltenegger, 2013; Sanromá *et al.*, 2014; O'Malley-James and Kaltenegger, 2018). Given the intricate relationship between geology and biology, it may not be possible for the entire surface of a planet to be composed of biological organisms. However, microorganisms can dominate oceans and greatly influence surface spectra of these environments. As long as these desert planets are water-limited and not water-less, it may be possible for microorganisms to cover their entire surface. For the purpose of this study, we assume optimal conditions where the planetary surface is dominated by one surface type and determine the upper limit of detection for these surface biosignatures. Assuming noise is dominated by non-planetary sources, Equation 1 shows that the detection times presented here are inversely proportional to the square of the fractional coverage of the surface biota. Future ecosystem-coupled general circulation models (*e.g.*, Krinner *et al.*, 2005; Sitch *et al.*, 2008; Harper *et al.*, 2018; Li *et al.*, 2019) may be able to provide simulation-derived constraints on microorganism surface coverage.

With these assumptions, required exposure times indicate that surface biosignatures could realistically be detected on Earth-like exoplanets using the HWO mission concept. Space telescopes, like Hubble, have been used to create images from hundreds of hours of exposure time. The Hubble Deep Field program collected continuous observations, resulting in 200 - 300 hours of exposure time, of a region in the northern Galactic hemisphere from December 18 - 30th, 1995 (Williams *et al.*, 1996). In 2004, the Hubble Ultra Deep Field, consisting of one million seconds of exposure time (~278 hours), found over 10,000 objects of a region in the southern sky (Beckwith *et al.*, 2006). Since 2004, images like those from Hubble Deep Field and Hubble Ultra Deep Field have also been combined to represent thousands of hours of exposure time (Whitaker *et al.*, 2019). Given how imaging missions have targeted a specific area of the sky for hundreds of hours of exposure time, the community could invest a similar amount of time to realistically detect pigment and photosynthetic red edge surface biosignatures.

*4.2 Effects of Cloud Conditions on Detection of Surface Biosignatures*

Cloud cover can introduce atmospheric effects, which can influence the detection of surface life (*e.g.,* Montañés-Rodríguez et al., 2006; Kaltenegger *et al.*, 2007; King *et al.*, 2013; O'Malley-James and Kaltenegger, 2018). Rayleigh scattering (size of particles is smaller than the wavelength of light) occurs during clear sky conditions, which ultimately contributes a blue color to the atmosphere. When cloud conditions increase, the Rayleigh scattering contribution is decreased and the Mie (size of particles is equal to the wavelength of light) and nonselective scattering (size of particles is larger than the wavelength of light) is increased, shifting overall reflectance to longer wavelengths and introducing greyness to the spectrum. Therefore, cloud cover can create relatively flattened or homogenous top-of-atmosphere reflectance spectra (*e.g.,* Montañés-Rodríguez et al., 2006; King *et al.*, 2013; Kreidberg *et al.*, 2014; Sing *et al.*, 2016; Diamond-Lowe *et al.*, 2018). As the reflectivity increases, the depth of spectral features in the planetary spectrum also diminish, including contributions from surface biosignatures (Kreidberg *et al.*, 2014; Sing *et al.*, 2016; Diamond-Lowe *et al.*, 2018; Kaltenegger *et al.*, 2007; King *et al.*, 2013; O'Malley-James and Kaltenegger, 2018). Our results show an increase in reflectivity with an increase in cloud cover (**Fig. 7**). Weakened surface features in the biotic spectrum reduces the difference between biotic and abiotic spectra, thereby increasing the time needed to distinguish between these spectra (in some cases; **Fig. 7**). As a result, surface biosignatures can become harder to detect with an increase in cloud coverage (Kaltenegger *et al.*, 2007; Tinetti *et al.,* 2006; O'Malley-James and Kaltenegger, 2018).

While an increase in cloud cover resulted in slower detection times for black and orange microbial mat, low to mid cloud cover resulted in shorter exposure times for *Arthrobacter* sp. and *Ectothiorhodospira* sp. str. BSL-9 (**Table 3; Fig. 9**). Exposure times for *Arthrobacter* sp. were fastest at 25% cloud cover, and exposure times for *Ectothiorhodospira* sp. str. BSL-9 were fastest at 50% cloud cover (**Table 3; Fig. 9**). For both *Arthrobacter* sp. and *Ectothiorhodospira* sp. str. BSL-9, the reduction in required exposure time gained by increasing the overall brightness of the planet was larger than the enhancement in exposure time that results from clouds reducing the contrast between the biotic and abiotic models. Pigment absorptions for both microbes were also low in reflectance in comparison with the red edge, which spanned low to mid reflectance values (**Fig. 7**). Therefore, an increase in reflectance could be improving the distinguishability between the biotic and abiotic spectrum of *Arthrobacter* sp. and *Ectothiorhodospira* sp. str. BSL-9, resulting in shorter exposure times. Low to mid cloud conditions ultimately have a substantial effect on the detection of surface biosignatures, sometimes improving exposure times for low reflective biosignatures attributed to certain microorganisms like our anoxygenic and nonphotosynthetic microbes.

*4.3 Effects of Atmospheric Water on Detection of Surface Biosignatures*

Atmospheric gaseous absorptions drastically affect surface biosignatures at longer wavelengths as opposed to shorter wavelengths. Water, oxygen, and ozone absorption features are present in top-of-atmosphere spectra beyond 0.58 μm and are located at wavelengths where phycobilins, chlorophylls, and the red edge are also absorbing and scattering light (**Fig. 5**). However, carotenoid pigments absorb light at shorter wavelengths and are thus not affected by water or other potential atmospheric absorption features (**Fig. 2**; **Fig. 5**). For this reason, carotenoids of both nonphotosynthetic and photosynthetic organisms that absorb at blue wavelengths may be more consistently detectable across a range of atmospheric compositions in comparison to surface biosignatures, such as chlorophyll or the red edge.

However, the detectability of photosynthetic biosignatures was better under low atmospheric water vapor conditions. Atmospheric absorptions in planetary spectra with Earth-like water vapor concentrations (**Fig. 5**) were deeper than their absorptions in low water vapor planetary spectra (**Fig. 8**). The reduction of these atmospheric absorptions, which are also located in similar wavelength regions as photosynthetic biosignatures, makes pigments and the red edge more visible in planetary spectra (**Fig. 8**). Additionally, exposure times required for detecting all photosynthetic organisms decreased between high and low atmospheric water vapor concentration so much that these organisms were ultimately faster to detect than nonphotosynthetic organisms. Therefore, Earth-like atmospheric water vapor concentrations could hinder the detection of photosynthetic biosignatures (Tinetti *et al.*, 2006). However, photosynthetic surface biosignatures in relatively drier atmospheres can potentially be more readily detectable than even nonphotosynthetic surface biosignatures.

*4.4 Implications for Next Generation Life Detection*

Our results support recommendations for future space-telescope designs. The relationship between the 1) exposure time, 2) distance to a planetary system, 3) telescope's inner working angle (IWA), and 4) longest wavelength observable using HWO determined the observational criteria necessary for the upper limit detection of surface biosignatures.

Most surface biosignatures required unreasonable detection times (>1,000 hours) beyond certain planetary distances (ranging from 9 - 15 pc) as a result of the IWA preventing spectral access to microbe-dependent pigment or red edge features at key wavelengths. The IWA determines the wavelength-dependent field-of-view of the telescope, where the inner "dark hole" provided by the coronagraph expands linearly with wavelength. For instance, the longest observable wavelength, or the maximum wavelength that could be detected in our telescope model with an IWA of 3.5 $\lambda/D$, became limited with an increase in planetary distance (**Fig. 10**). Therefore, surface biosignatures located at shorter wavelengths had shorter exposure times and were more detectable than surface biosignatures that were present at longer wavelengths. When the IWA was decreased from 3.5 $\lambda/D$ to 2.5 $\lambda/D$, all microorganisms, with the exception of *Arthrobacter* sp., had shorter exposure times at further distances, owing to improved spectral access to important redder-wavelength features and allowing surface biosignatures present at longer wavelengths to be detectable at further planetary distances (**Fig. 10**). Because our microbial spectra were easier to detect at further distances with an IWA of 2.5 $\lambda/D$ relative to 3.5 $\lambda/D$ (**Fig. 10**), we demonstrate that a large IWA could exclude planetary targets from a telescope's field-of-view at key red/near-infrared wavelengths as distance from the planet increases. Thus, a tightest possible IWA should be designed for future space-based telescopes for planets over 5 pc away.

Even with a larger IWA of 3.5 $\lambda/D$, distinguishing pigments and the photosynthetic red edge could be possible for a number of rocky exoplanets within at least 11 pc of Earth. The likelihood of detecting the red edge however, decreases with an increase in planetary distance because the longest observable wavelength becomes restricted to shorter wavelengths (**Fig. 10**). In contrast, nonphotosynthetic and photosynthetic carotenoid pigments will continue to be detectable at further distances given their presence at shorter wavelengths (**Fig. 2**; **Fig. 4**). Thus, targeting the detection of chlorophylls and the red edge may only be useful for closer planets, whereas carotenoid pigments could be a beneficial surface biosignature to target for planets beyond 11 pc.

When considering life on other worlds however, the evolution of photosynthesis could be different from what is seen on Earth (Tinetti *et al.*, 2006; Kiang *et al.* 2007a; Kiang *et al.*, 2007b). Photosynthetic pigments tend to harvest light at the maximum photon flux emitted by the stellar host that is within the atmospheric window, or region of the electromagnetic spectrum where energy can be transmitted through the atmosphere (Kiang *et al.* 2007a; Kiang *et al.*, 2007b). The maximum photon flux varies for each star type based on Planck's blackbody function. For planets orbiting other stellar hosts, photosynthetic pigments and the red edge could shift to longer wavelengths when orbiting a low temperature star and shorter wavelengths if orbiting a high temperature star (Kiang *et al.* 2007a; Kiang *et al.*, 2007b). If observing an Earth-like exoplanet orbiting a high temperature star, any potential photosynthetic red edge would be detectable across a wide range of planetary distances even with a larger IWA because it would be present at shorter wavelengths. Additionally, the photosynthetic red edge could be more distinguishable than false positive mineralogies because the mineral features could remain at the same wavelength, whereas possibly the red edge could shift to much shorter wavelengths (Tinetti *et al.*, 2006). In contrast, nonphotosynthetic pigments could be difficult to predict with stellar host variation as they are not functionally tied to the photon flux, indicating that a caveat to detecting nonphotosynthetic pigments would be verifying absorption features are a result of life as we know it (Schwieterman *et al.*, 2015). In this instance, the predictability of photosynthetic biosignatures, like the red edge, for a range of stellar hosts and planetary distances could make these surface biosignatures more reliable to detect on Earth-like exoplanets than nonphotosynthetic biosignatures.

Given the wide wavelength range of pigment absorptions within *Ectothiorhodospira* sp. str. BSL-9, our anoxygenic photosynthetic microbe was fastest to detect compared to our oxygenic photosynthetic and nonphotosynthetic microbial communities. Our anoxygenic photosynthetic microbe absorbing wavelengths of light across both blue and red wavelengths created a strong biotic signature no other organism could create. While oxygenic photosynthetic microbes, including black and orange microbial mat, also have pigment absorption features across both blue and red wavelengths (**Fig. 2**), a hindrance for detecting these communities was an iron oxide false positive analog for the red edge. Both black and orange microbial mat had slower exposure times as a direct result of the red edge being fairly similar to the iron oxide false positive. Therefore, *Ectothiorhodospira* sp. str. BSL-9 benefited from the lack of similarity against a false positive analog. While an increase in spectral slope for some anoxygenic photosynthetic microbes around 1 μm, called a NIR "edge," could be mistaken for an iron mineral slope (Kiang *et al.*, 2007b; Sanromá *et al.*, 2014; Schwieterman *et al.*, 2018), the wavelength range of a future telescope may end at 1 μm, cutting off the NIR "edge." Therefore, even this potential false positive analog may not be reliable to compare against all anoxygenic photosynthetic life.

Our results demonstrate that when thinking about the types of surface life that could be detected on an Earth-like exoplanet, organisms with many different types of surface biosignatures could be most beneficial to model. Planetary distance will pose a challenge to the detection of the red edge, assuming a similar star type as the Sun, and therefore, being able to detect carotenoids in blue wavelength regions will be important when the red edge cannot be detected. Additionally, the ability to detect multiple types of biosignatures strengthens the possibility for a positive detection of life on another world. For instance, detecting carotenoids, phycobilins, chlorophylls, and in some cases a red edge could result in more robust detections of surface life on an Earth-like exoplanet. Therefore, continued exploration of the spectral characteristics for many diverse organisms and ecosystems is necessary in further identifying a range of potentially detectable surface biosignatures.

**Conclusion**

We compared the detectability of surface biosignatures affiliated with three different extremophiles: 1) oxygenic, 2) anoxygenic, and 3) nonphotosynthetic desert microorganisms. We used a 1D atmospheric radiative transfer model and a HWO instrument noise model to determine required exposure times between all surface biosignatures. We assumed global coverage of microorganisms to achieve optimal detection and adjusted atmospheric water vapor to be that of Earth's and 10% of Earth's for a desert-like world. For Earth-like water vapor planets, we also modeled planetary spectra with 0%, 25%, 50%, and 75% cloud cover. We also explored varying inner working angles of the HWO model to determine distance and longest observable wavelength constraints on the detection of biosignatures for 0% cloud cover for Earth-like and desert-like atmospheric water vapor planets and 50% cloud cover for Earth-like atmospheric water vapor planets. We found that low to mid cloud coverage aided in faster detection times for our anoxygenic and nonphotosynthetic microbes, and nonphotosynthetic pigments were unaffected by atmospheric changes in water vapor. However, our anoxygenic photosynthetic microbe was fastest to detect as a result of a lack of reliable false positive analogs and the presence of multiple surface biosignatures that spanned a wide wavelength range. For modeled planets with 0% cloud coverage, we found that detections of pigments and the photosynthetic red edge could be realistically achieved with exposure times less than 500 hr (50 hr) at distances within 15 pc (11 pc) with an inner working angle of 3.5 $\lambda/D$. A tighter telescope inner working angle of 2.5 $\lambda/D$ enabled planetary targets at further distances to remain in view, enabling the detection of surface biosignatures at further distances. However, as planetary distance increases, the likelihood of detecting the red edge decreases because the inner working angle restricts observations to short wavelengths. As a result, the red edge could be detectable even against false positives on planets at smaller distances (<11 pc), whereas carotenoid pigments are more reliable surface biosignatures for detecting on planets further away (>11 pc). Our findings suggest that given their lack of false positives and presence at shorter wavelengths, carotenoids are discriminating biosignatures useful for life detection studies. Future work to comprehensively study the biological and spectral relationship of carotenoids will inform studies such as this. Additionally, retrieval models with a parameterized surface albedo spectrum (*e.g.*, Gomez Barrientos *et al.*, 2023) could be applied to cases with various surface coverage fractions of organisms with pigments to more accurately model reflectance spectra of a diversity of potential ecosystems on exoplanets. Altering the stellar type and planetary size in these models will also improve our understanding of a range of habitability limitations on biosignature detection.

**Acknowledgements**

The authors would like to thank Prof. Mark Loeffler for his thorough suggested edits on the analysis, structure, figures, and writing of the manuscript. We would also like to thank Prof. Andrew Richardson and his 2022 Scientific Paper Writing Seminar Class for their tireless feedback on this manuscript. Special thanks to Drs. Lori Pigue and Christian Tai Udovicic for the amount of guidance they provided on the manuscript and figures. Thank you also to members of the Habitability, Atmospheres, and Biosignatures lab for their help with this project. In particular, thank you to Dr. Chris Wolfe, who helped us use Northern Arizona University's Monsoon computing cluster for simulating our SMART spectra. Also, thank you to PhD candidate, James Windsor, for his helpful comments and discussion of the project and its figures. Lastly, thank you to Drs. Amber Young and Arnaud Salvador for providing feedback on figures.

**Data Availability Statement**

Spectral data used in this manuscript was obtained from the following databases: Hegde *et al.* (2015), Kokaly *et al.* (2017), and Borges *et al.* (2023). The HWO telescope model was modified from Lustig-Yaeger's coronagraph python script affiliated with Lustig-Yaeger *et al.* (2019) and is available on GitHub at: https://github.com/jlustigy/coronagraph.git. All other python scripts used to generate data in this manuscript are available on GitHub at this link: https://github.com/borgess28/Detectability-of-Surface-Biosignatures-for-Directly-Imaged-Rocky-Exoplanets.git.

**Authorship Contribution Statement**

**S.R.B**: Conceptualization (lead); data curation (lead); methodology (equal); project administration (lead); software (supporting); supervision (equal); visualization (lead); writing – original draft (lead); writing – review and editing (lead). **G.G.J**: Conceptualization (supporting); formal analysis (lead); methodology (equal); software (lead); writing – review and editing (supporting). **T.D.R**: Conceptualization (supporting); methodology (equal); project administration (supporting); resources (lead); supervision (equal); validation (lead); visualization (supporting); writing – review and editing (supporting).

**Author Disclosure Statement**

No competing financial interests exist.

**Funding Statement**

This research was supported by the National Science Foundation Graduate Research Fellowship awarded to S.R.B. and NASA/NAU Space Grant awarded to G.G.J. T.D.R. gratefully acknowledges support from NASA's Exoplanets Research Program (No.~80NSSC18K0349), Exobiology Program (No.~80NSSC19K0473), Habitable Worlds Program (No.~80NSSC20K0226), the Nexus for Exoplanet System Science Virtual Planetary Laboratory (No.~80NSSC18K0829), and the Cottrell Scholar Program administered by the Research Corporation for Science Advancement. Hyperspectral data of black and orange microbial mat from Antarctica were collected as a part of [Borges *et al.* (2023)](#) under the U.S. National Science Foundation Office of Polar Programs awards 1758224 (Antarctic Earth Science), 1745053 (Antarctic Organisms & Ecosystems), and 1744849 (Antarctic Organisms & Ecosystems).

Assistance with data management and archival for this dataset was provided through the McMurdo Long-Term Ecological Research grants #OPP-1637708 and #OPP-2224760.

10.1016/j.chemphys.2008.12.026.

**Tables**

| Instrument Model | | | | | | |
|---|---|---|---|---|---|---|
| | **Biotic Spectrum (X model)** | **Abiotic Spectrum (Y model)** | **Cloud coverage** | | | |
| **Earth atmospheric water vapor** | Black microbial mat | Black microbial mat | 0% | 25% | 50% | 75% |
| | Orange microbial mat | Orange microbial mat | 0% | 25% | 50% | 75% |
| | *Arthrobacter* sp. | *Arthrobacter* sp. | 0% | 25% | 50% | 75% |
| | *Ectothiorhodospira* sp. str. BSL-9 | *Ectothiorhodospira* sp. str. BSL-9 | 0% | 25% | 50% | 75% |
| **10% of Earth atmospheric water vapor** | Black microbial mat | Black microbial mat | 0% | | | |
| | Orange microbial mat | Orange microbial mat | 0% | | | |
| | *Arthrobacter* sp. | *Arthrobacter* sp. | 0% | | | |
| | *Ectothiorhodospira* sp. str. BSL-9 | *Ectothiorhodospira* sp. str. BSL-9 | 0% | | | |

**Table 1:** Top-of-atmosphere reflectance spectra generated for each experimental treatment input into the instrument noise model.

| HWO Telescope Parameters | | |
|---|---|---|
| Parameter | Description | Value |
| D | telescope diameter | 6 m |
| C | coronagraph design contrast | $1.0 \times 10^{-10}$ |
| $A_{collect}$ | mirror collecting area | 36 m$^2$ |
| $D_{inscribed}$ | inscribed telescope diameter used for lenslet calculations | 6 m |
| $T_{sys}$ | telescope temperature | 270 K |
| $\theta_{OWA}$ | Outer Working Angle (λ/D) | 24 |
| $q$ * charge transfer term | detector quantum efficiency * charge transfer term | 0.9 * 0.75 |
| $\epsilon_{sys}$ | telescope emissivity | 0.7 |
| $\Omega$ | photometry aperture size as a multiple of $\lambda/D$ | 0.61 |
| $\Delta t_{max}$ | Maximum exposure time | $1.0 \times 10^3/3600$ hr |
| | **Visible channel** | |
| $\theta_{IWA}$ | Inner Working Angle ($\lambda/D$) | 3.5 & 2.5 |
| R | instrument spectral resolution | 140 |
| T | telescope throughput | 0.15 |
| $D_e^-$ | dark current | $3.0 \times 10^{-5}$ s$^{-1}$ |
| $R_e^-$ | read noise counts per pixel | 0 |
| $\lambda_{max}$ | maximum wavelength | 1.03 μm |
| $\lambda_{min}$ | minimum wavelength | 0.4 μm |
| $R_c$ | clock induced charge (counts/pixel/photon) | $2.0 \times 10^{-03}$ |

**Table 2:** HWO telescope parameters used in our instrument model.

| | Required Exposure Time (hr) | | | |
|---|---|---|---|---|
| **Cloud Cover (%)** | **0** | **25** | **50** | **75** |
| **Black mat** | 130 | 140 | 210 | 460 |
| **Orange mat** | 49 | 48 | 58 | 94 |
| *Arthrobacter* sp. | 94 | 71 | 75 | 100 |
| *Ectothiorhodospira* sp. str. BSL-9 | 41 | 10 | 7.5 | 10 |

**Table 3:** Exposure time needed to achieve a S/N ratio of 5 at 5 pc with an inner working angle of 3.5 $\lambda/D$ for detecting black microbial mat, orange microbial mat, *Arthrobacter* sp., and *Ectothiorhodospira* sp. str. BSL-9 with 0%, 25%, 50%, and 75% cloud cover at Earth-like atmospheric water vapor concentrations.

| Required Exposure Time (hr) | |
|---|---|
| **Black mat** | 73 |
| **Orange mat** | 39 |
| *Arthrobacter* sp. | 93 |

| *Ectothiorhodospira* sp. str. BSL-9 | 31 |

**Table 4:** Exposure time needed to achieve a S/N ratio of 5 at 5 pc with an inner working angle of 3.5 λ/*D* for detecting black microbial mat, orange microbial mat, *Arthrobacter* sp., and *Ectothiorhodospira* sp. str. BSL-9 with low atmospheric water vapor concentration and no cloud cover.

**Figures**

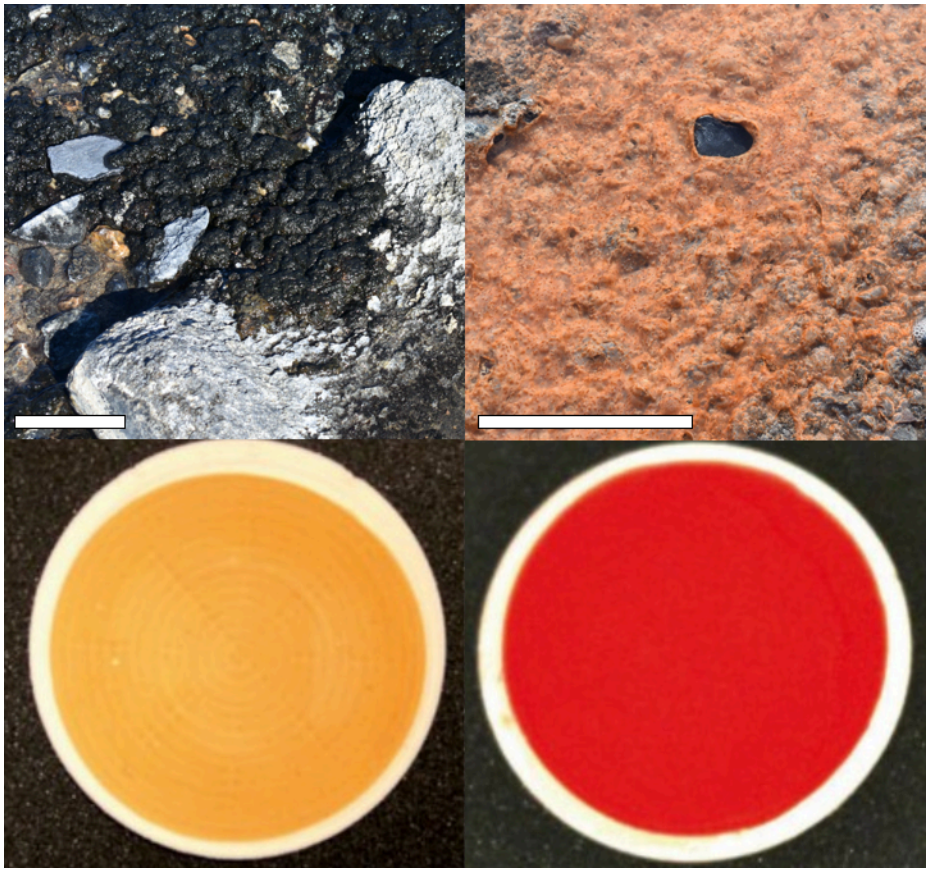

**Figure 1:** Images of the four microbes used in this study. Black and orange microbial mats (top left and right, respectively) are oxygenic photosynthetic microbial communities from the Fryxell Basin of Taylor Valley, Antarctica. Scale bars in both images are 10 cm. Both spectra and images were collected in the field as part of the database from Borges *et al.* (2023). *Arthrobacter* sp. and *Ectothiorhodospira* sp. str. BSL-9 (bottom left and right, respectively) are nonphotosynthetic and anoxygenic photosynthetic microbes grown on 2.5 cm diameter filters in the lab as part of work conducted in Hegde *et al.* (2015). Images and spectra are from Hegde *et al.* (2015).

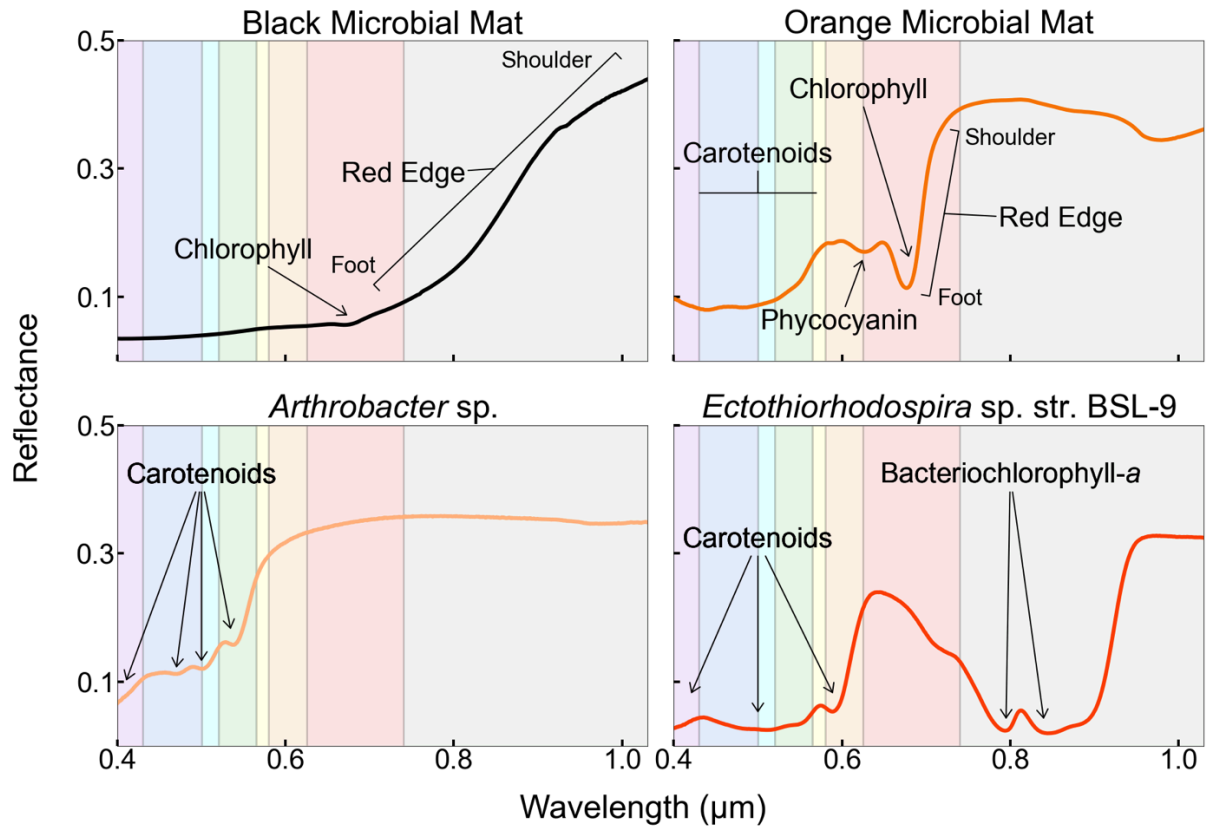

**Figure 2:** Reflectance spectra of all four microbes with biosignatures and wavelength regions annotated. Colors represent regions of the visible wavelength range, and the gray area represents the near-infrared region of the electromagnetic spectrum. Annotated biosignatures for black microbial mat include a chlorophyll absorption feature and a photosynthetic red edge. Orange microbial mat also has a photosynthetic red edge in addition to chlorophyll, phycocyanin, and carotenoid absorptions. Biosignatures within the *Arthrobacter* sp. spectrum include carotenoids,

and *Ectothiorhodospira* sp. str. BSL-9 has both carotenoid and bacteriochlorophyll-*a* absorptions.

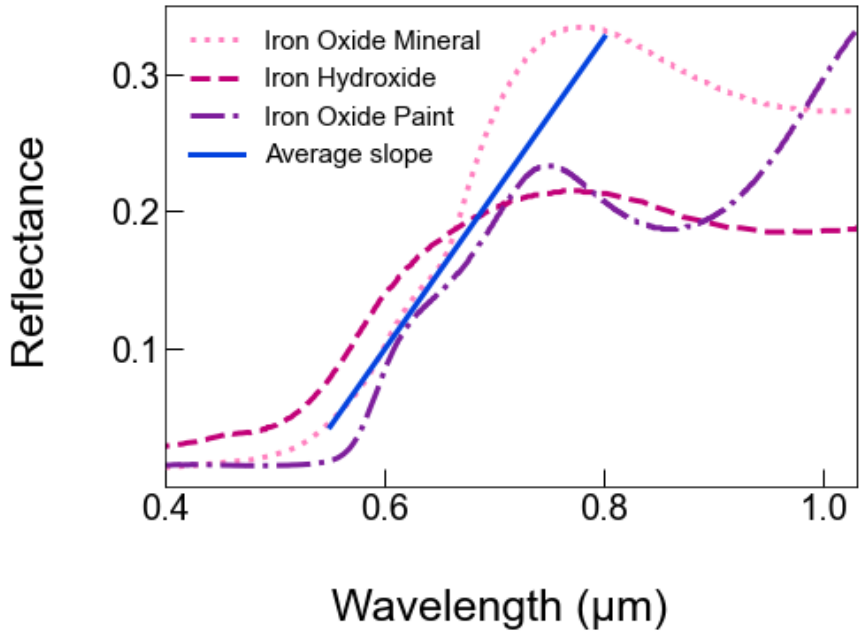

**Figure 3:** Iron oxide and hydroxide spectra from Kokaly *et al.* (2017) plotted against their average slope between 0.6 – 0.8 μm. The iron oxide mineral is Iron_Oxide SA-480665 gamma ASDHRa AREF (speclib07a rec 18816) from Kokaly *et al.* (2017)'s database and is a powder purchased from Sigma Aldrich chemistry. Iron Hydroxide (Fe-Hydroxide SU93-106 amorph BECKb AREF (speclib07a rec=12451)) is the coating of a river cobble which was collected in Alamosa River, CO, USA 15 km downstream of a mine that released metal-rich mine waste water into the river (Kokaly *et al.*, 2017). Iron Oxide Paint refers to Iron oxide_#4820 GDS782 ASDFRa AREF (speclib07a rec=18798), which is a Kremer Pigments Inc. powdered paint pigment from Kokaly *et al.* (2017).

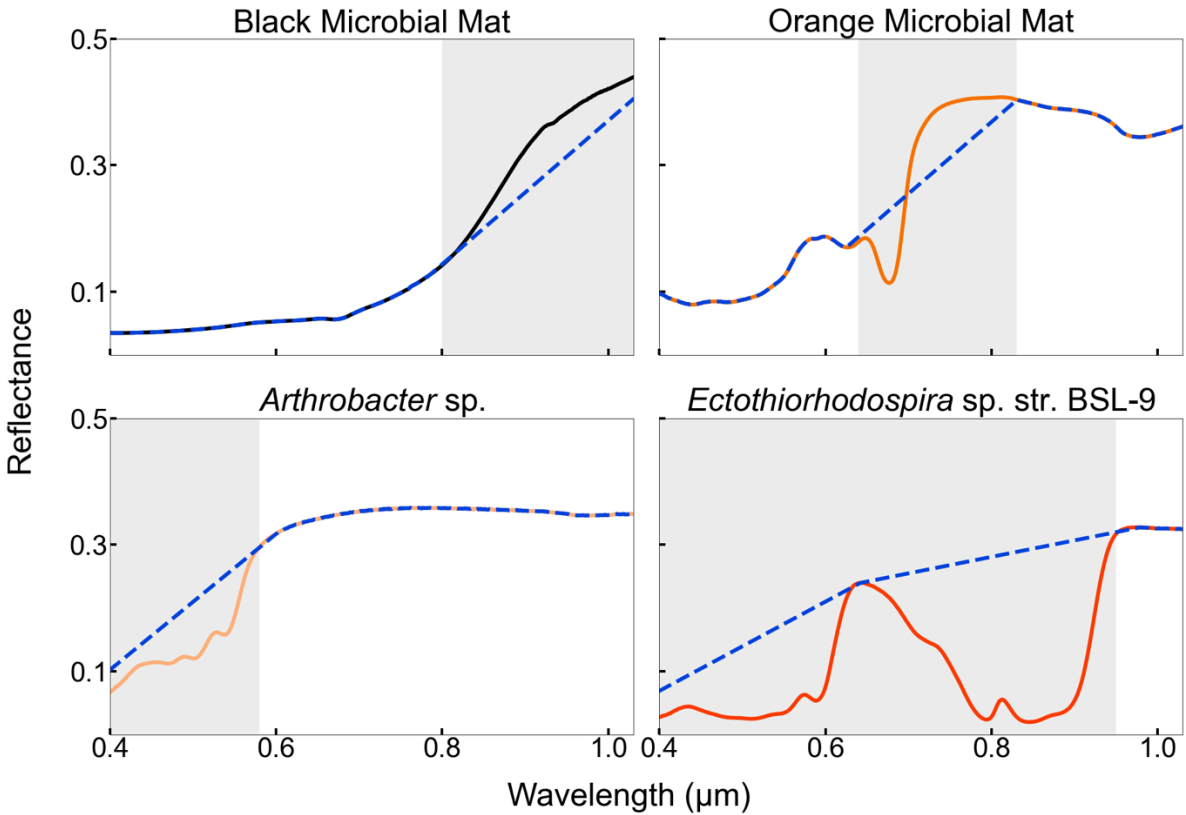

**Figure 4:** Biotic (solid) and abiotic (dashed) reflectance spectra for each surface microbial community. Biosignatures were removed from measured microbial spectra to create abiotic spectra, and the photosynthetic red edge was replaced with the averaged iron oxide "red edge" slope. Grey regions indicate differences between biotic and abiotic spectra, which coincide with the location of the surface biosignatures of interest.

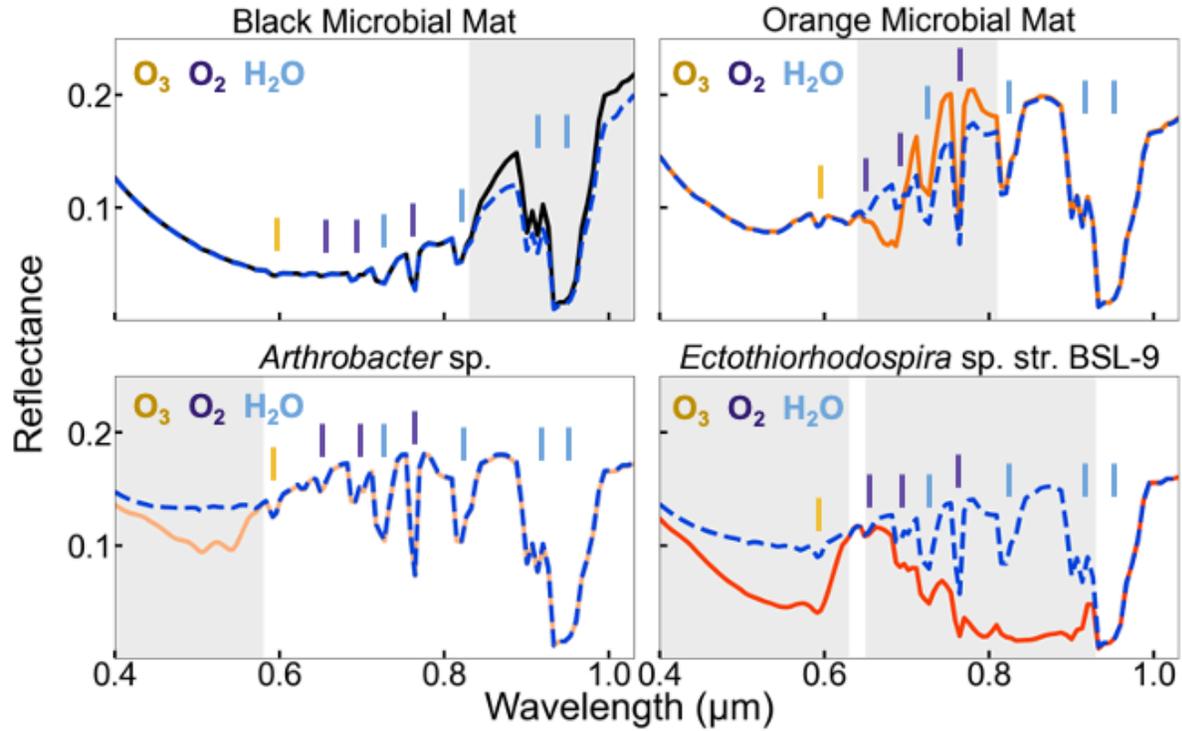

**Figure 5:** Top-of-atmosphere reflectance spectra at clear sky conditions for both biotic (solid) and abiotic (dashed) spectra of each microbial community. Atmospheric gas absorption features are depicted, and grey regions indicate areas where surface biosignatures affiliated with each microbial community are influencing the overall top-of-atmosphere reflectance.

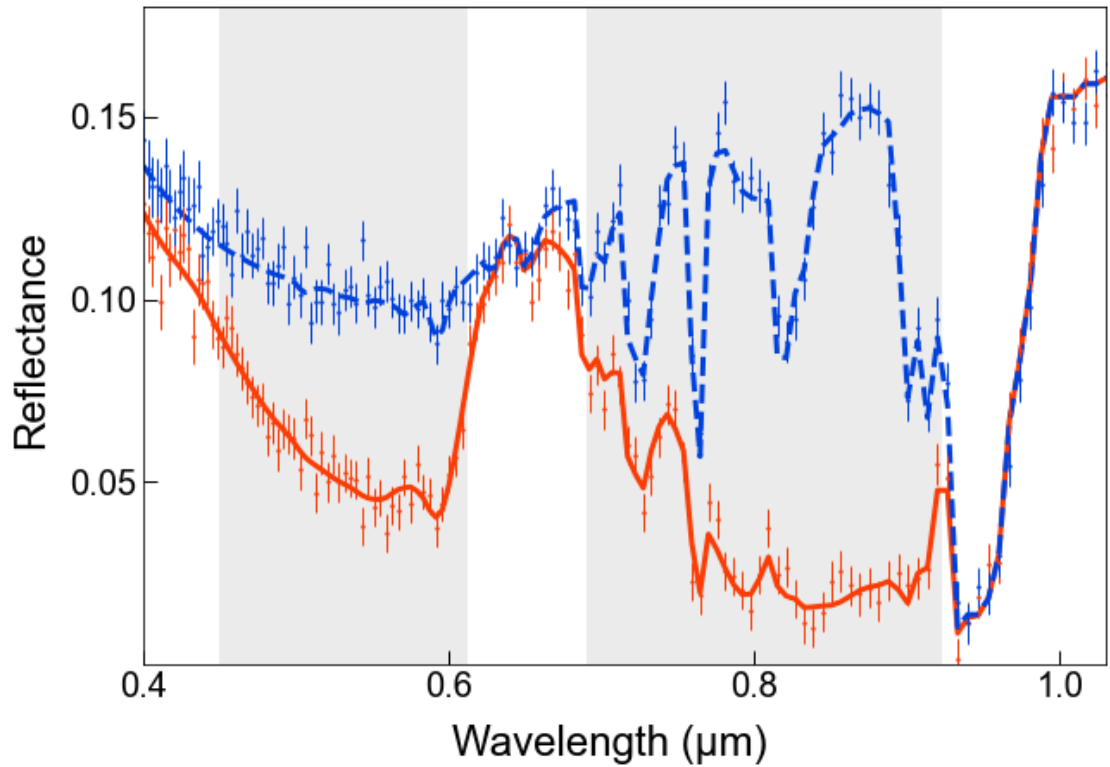

**Figure 6:** Synthetic telescope observational spectra of biotic (solid line) and abiotic (dashed line) *Ectothiorhodospira* sp. str. BSL-9 cases without clouds at 7.5 pc and 100 hours of exposure time. Error bars for both spectra were calculated by first determining the S/N at each wavelength value and then dividing the observed spectrum by these S/N values. Grey areas indicate the wavelength region where error bars do not overlap and biologic features are influencing the biotic spectrum.

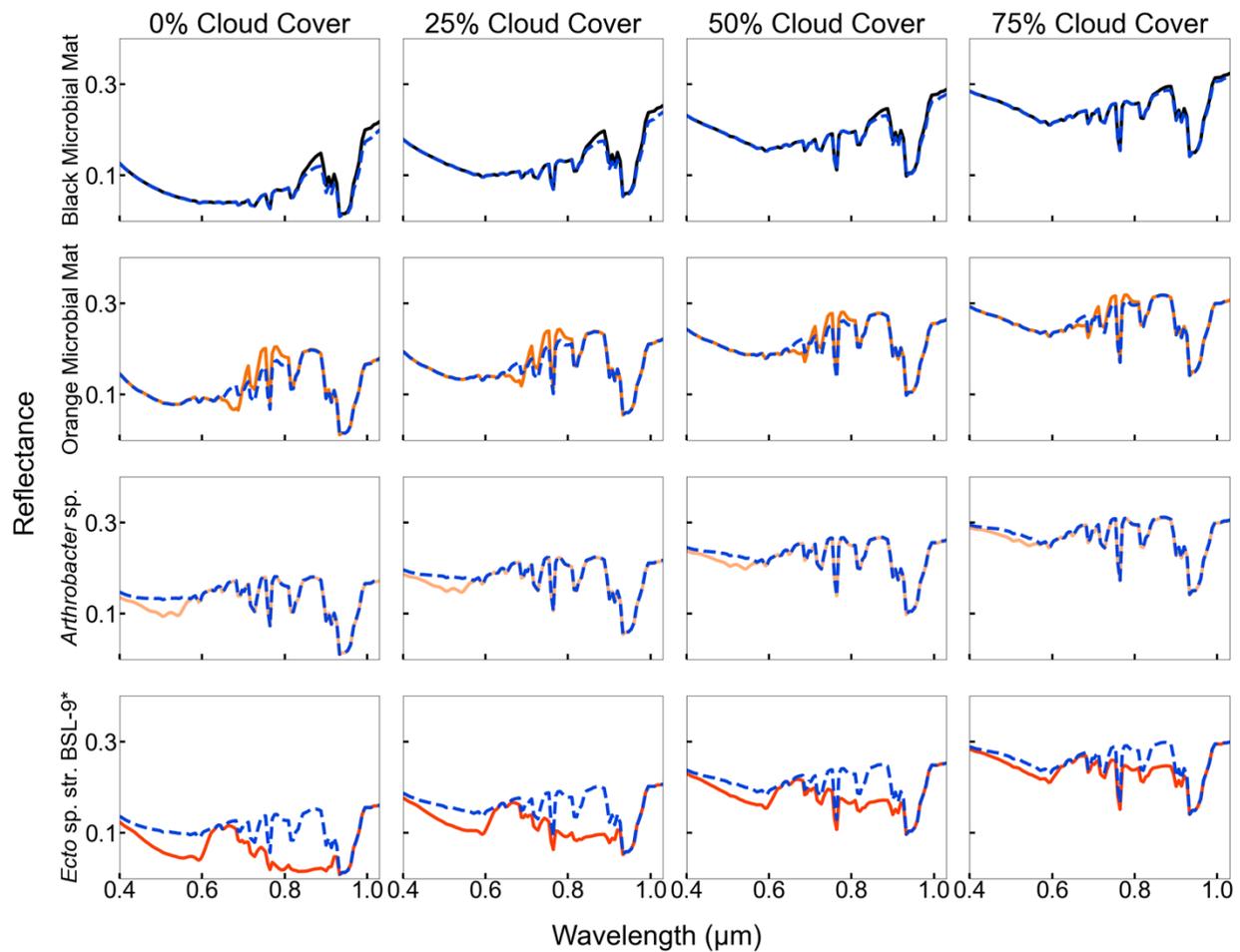

**Figure 7:** Top-of-atmosphere reflectance spectra with 0%, 25%, 50%, and 75% cloud cover for both biotic (solid) and abiotic (dashed) spectra of each microbial community. Top row: black microbial mat. Second row: orange microbial mat. Third row: *Arthrobacter* sp. Last row: *Ectothiorhodospira* sp. str. BSL-9 (name shortened in plot). Overall reflectance of all spectra increases with increasing cloud cover.

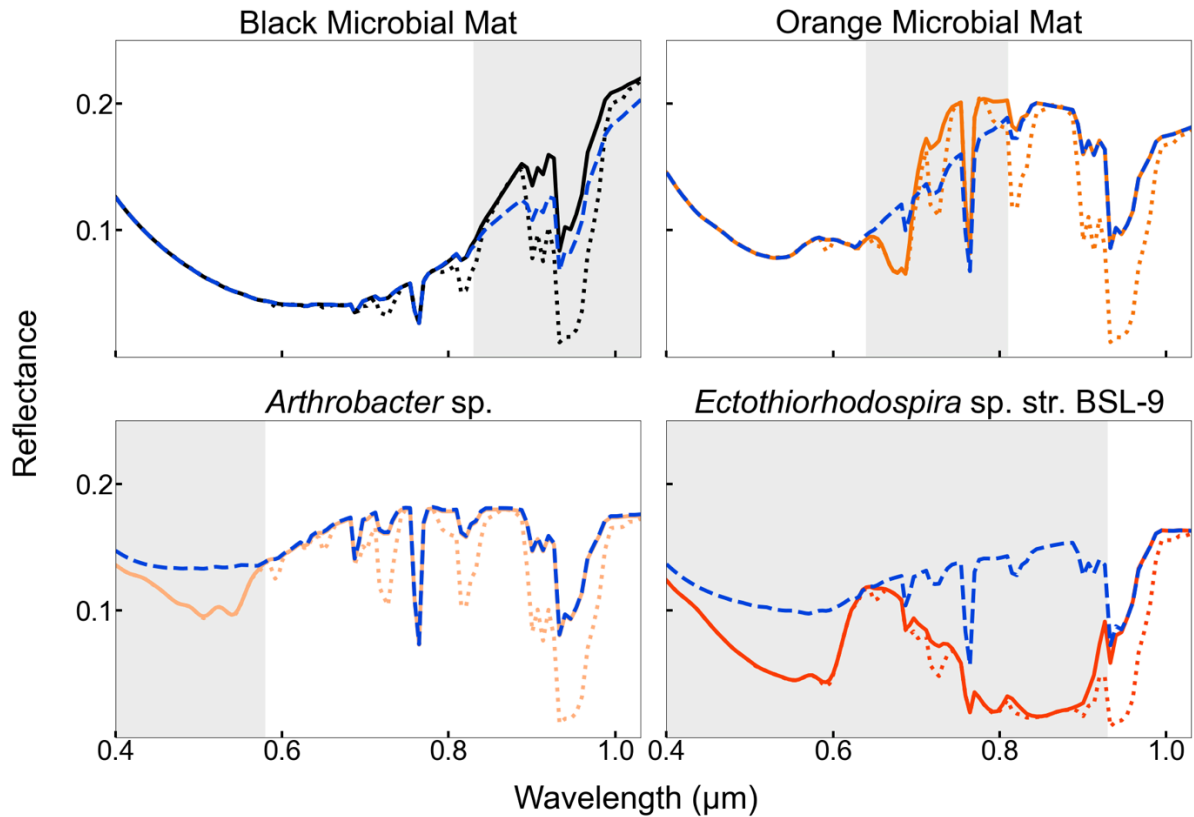

**Figure 8:** Top-of-atmosphere reflectance spectra at clear sky conditions with low atmospheric water vapor conditions (solid) and Earth-like atmospheric water vapor concentrations (dotted) for black microbial mat, orange microbial mat, *Arthrobacter* sp., and *Ectothiorhodospira* sp. str. BSL-9. Abiotic top-of-atmosphere spectra with low atmospheric water vapor and clear skies is also shown for each microbe (dashed). Grey regions represent areas of the spectrum influenced by surface biosignatures.

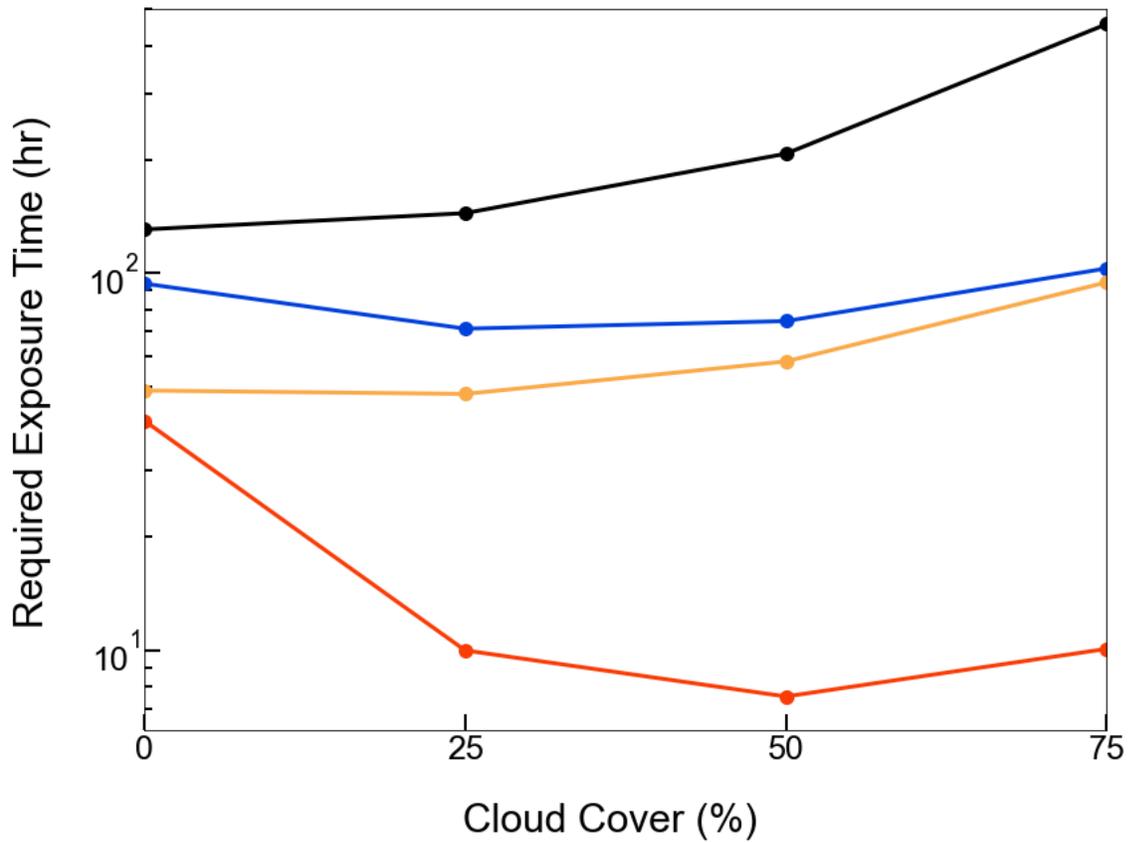

**Figure 9:** Required exposure time for detecting black microbial mat (black), orange microbial mat (orange), *Arthrobacter* sp. (blue), and *Ectothiorhodospira* sp. str. BSL-9 (red) with Earth-like atmospheric water vapor conditions at each cloud cover percentage, a planetary distance of 5 pc, and an inner working angle of 3.5 λ/*D*.

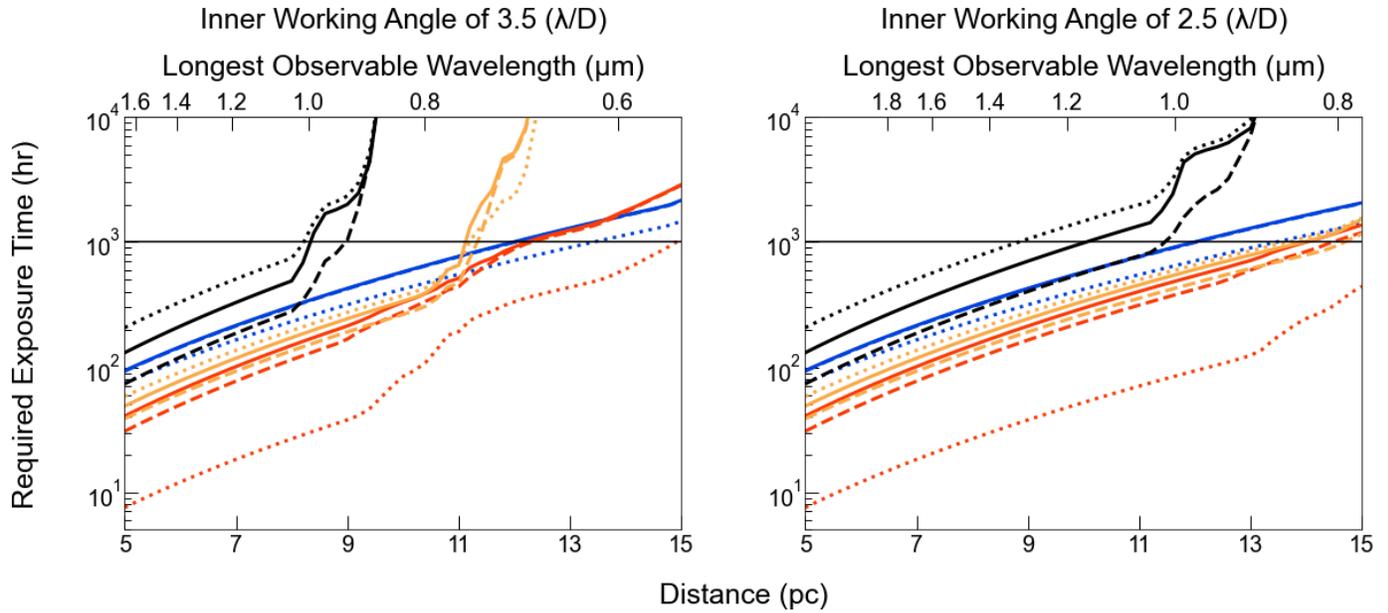

**Figure 10:** Required exposure time versus distance and longest observable wavelength with an inner working angle of 3.5 $\lambda/D$ (left) and 2.5 $\lambda/D$ (right) for black microbial mat (black), orange microbial mat (orange), *Arthrobacter* sp. (blue), and *Ectothiorhodospira* sp. str. BSL-9 (red). 0% cloud cover (solid) and 50% cloud cover (dotted) with an Earth-like amount of atmospheric water vapor is plotted against 10% of Earth's atmospheric water vapor at 0% cloud cover (dashed). The black horizontal line at 1,000 hours represents a maximum reasonable exposure time for detecting each microbial community.